\documentclass[12pt, a4paper]{article}      
\usepackage[dvips]{graphicx}
\usepackage{wrapfig}
\usepackage{amssymb}
\usepackage{amsmath}
\usepackage[a4paper,hmargin={3cm,.9in},height=10in]{geometry}
\usepackage{mathrsfs}
\usepackage{eurosym}
\usepackage{dsfont}
\begin{document}

\title{Quantum
information approach\\ to the ultimatum game
}


\author{Piotr Fr\c{a}ckiewicz}
\newtheorem{lemma}{Lemma}
\newtheorem{definition}{Definition}[section]
\newtheorem{theorem}{Theorem}
\newtheorem{proposition}{Proposition}[section]
\newtheorem{example}[proposition]{Example}
\newenvironment{xx}{\noindent\textit{Proof.}}
{\nolinebreak[4]\hfill$\square$\\\par}

\author{\textsc{Piotr Fr\c{a}ckiewicz} \\Institute of Mathematics, Polish Academy of Sciences\\
00-956 Warsaw, Poland}

\maketitle

\begin{abstract}
\noindent The paper is devoted to quantization of extensive games
with the use of both the Marinatto-Weber and the
Eisert-Wilkens-Lewenstein concept of quantum game. We revise the
current conception of quantum ultimatum game and we show why the
proposal is unacceptable. To support our comment, we
  present the new idea of the quantum ultimatum game. Our scheme also makes a point of departure for a~protocol to quantize extensive
  games.
\end{abstract}

\section{Introduction}
During the last twelve years of research into quantum games the
theory has been already extended beyond $2\times2$ games. Since
majority of noncooperative conflict problems are described by
games in extensive form, it is interesting to place extensive
games in the quantum domain. Although there is still no commonly
accepted idea of how to play quantum extensive games, we have
proved in \cite{fracor3} that it is possible to use the framework
\cite{eisert2} of strategic quantum game to get some insight into
quantum extensive games. Namely, we have shown that a~Hilbert
space $\mathscr{H} = \mathds{C}^2 \otimes \mathds{C}^2 \otimes
\mathds{C}^2$, a unit vector $|\psi_{\mathrm{in}}\rangle \in
\mathscr{H}$, the collection of subsets
$\{\mathcal{U}_{j}\}_{j=1,2,3}$ of $\mathsf{SU}(2)$, and
appropriately defined functionals $E_{1}$ and $E_{2}$ express the
normal representation of a two stage sequential game. Moreover, it
allows to get a result inaccessible in the game played
classically. In this paper, the above-mentioned quantum computing
description will be used to the two proposed variants of the
ultimatum game \cite{guth}. It is a game in which two players take
part. The first player proposes one of two proposals how to divide
a fixed amount of good. Then the second player either accepts or
rejects the proposal. In the first case, each player receives the
part of goods according to player 1's proposal. In the second
case, the players receive nothing. A game-theoretic analysis shows
that player~1 is in a better position. Since player 2's rational
move is to accept each proposal, player 1's rational move is to
make the best proposal for her. As we will show in this article,
the Eisert-Wilkens-Lewenstein (EWL) approach \cite{eisert} as well
as Marinatto-Weber (MW) approach \cite{marinatto} can change the
scenario of the ultimatum game significantly improving the
strategic position of player 2. Our paper also provides an
argument indicating
 that the previous idea \cite{mendes} of quantum ultimatum game is not sufficient to describe the game in the quantum
 domain. We will explain that, in fact, the formerly proposed protocol does not quantize the ultimatum game but another $2\times2$
 game. The last part of the paper is devoted to a form of a game tree
 where we provide the procedure how to determine the game tree when the game is played according to the MW
 approach.

\section{Preliminaries to game theory}
Definitions in the preliminaries are based on \cite{osborne}. This
section starts with a definition of a~finite extensive game.
\begin{definition} Let the following components be given.
\begin{itemize}
\item A finite set $N = \{1,2,\dots,n\}$ of players. \item A set
$H$ of finite sequences that satisfies the following two
properties:
\begin{enumerate}
\item the empty sequence $\emptyset$ is a member of $H$; \item if
$(a_k)_{k = 1,2,\dots, K} \in H$ and $K>1$ then $(a_k)_{k =
1,2,\dots, K-1} \in H$.
\end{enumerate}
Each member of $H$ is a history and each component of a history is
an action taken by a player. A history $(a_{1}, a_{2},\dots,
a_{K}) \in H$ is terminal if there is no $a_{K+1}$ such that
$(a_{1}, a_{2},\dots, a_{K}, a_{K+1}) \in H$. The set of actions
available after the nonterminal history $h$ is denoted $A(h) = \{a
\colon (h,a) \in H\}$ and the set of terminal histories is denoted
$Z$. \item The player function $P \colon H \setminus  Z
\rightarrow N \cup \{c\}$ that points to a player who takes an
action after the history $h$. If $P(h) = c$ then chance (the
chance-mover) determines the action taken after the history $h$.
\item A function $f$ that associates with each history $h$ for
which $P(h) = c$ an~independent probability distribution
$f(\cdot|h)$ on $A(h)$.\item For each player $i\in N$ a partition
$\mathcal{I}_{i}$ of $\{h \in H \setminus Z: P(h) = i\}$ with the
property that for each $I_{i} \in \mathcal{I}_{i}$ and for each
$h$, $h'$ $\in I_{i}$  an equality $A(h) = A(h')$ is fulfilled.
Every information set $I_{i}$ of the partition corresponds to the
state of player's knowledge. When the player makes move after
certain history $h$ belonging to $I_{i}$, she knows that the
course of events of the game takes the form of one of histories
being part of this information set. She does not know, however, if
it is the history $h$ or the other history from $I_{i}$. \item For
each player $i \in N$ a utility function $u_{i}\colon Z \to
\mathds{R}$ which assigns a number (payoff) to each of the
terminal histories.
\end{itemize}
A six-tuple $\left(N, H, P, f, \{\mathcal{I}_{i}\},
\{u_{i}\}\right)$ is called a finite extensive game.
\label{edefinition}
\end{definition}
Our deliberations focus on games with perfect recall (although
Def.~\ref{edefinition} defines extensive games with imperfect
recall as well) - this means games in which at each stage every
player remembers all the information about a course of the game
that she knew earlier (see \cite{myerson} and \cite{osborne} to
learn about formal description of this feature).

The notions: action and strategy mean the same in static games,
because the players choose their actions once and simultaneously.
In the majority of extensive games a player can make her decision
about an action depending on all the actions taken previously by
herself and also by all the other players. In other words, players
can make some plans of actions at their disposal such that these
plans point out to a specific action depending on the course of a
game. Such a plan is defined as a strategy in an extensive game.
\begin{definition}
A pure strategy $s_{i}$ of a player $i$ in a game $(N, H, P,
f_{c}, \{\mathcal{I}_{i}\}, \{u_{i}\})$ is a~function that assigns
an action in $A(I_{i})$ to each information set $I_{i} \in
\mathcal{I}$. \label{strategy}
\end{definition}
Like in the theory of strategic games, {\em a mixed strategy}
$t_{i}$ of a player $i$ in an extensive game is a probability
distribution over the set of player $i$'s pure strategies.
Therefore, pure strategies are of course special cases of mixed
strategies and from this place whenever we shall write {\em
strategy} without specifying that it is either pure or mixed, this
term will cover both cases. Let us define an {\em outcome $O(s)$}
of a strategy profile $s = (s_{1}, s_{2},\dots, s_{n})$ in an
extensive game without chance moves to be a terminal history that
results when each player $i \in N$ follows the plan of $s_{i}$.
More formally, $O(s)$ is the history $(a_{1}, a_{2},\dots, a_{K})
\in Z$ such that for $0 \leq k < K$ we have $s_{P(a_{1},
a_{2},\dots, a_{k})}(a_{1}, a_{2},\dots, a_{k}) = a_{k+1}$. If $s$
implies a history that contains chance moves, the outcome $O(s)$
is an appropriate probability distribution over histories
generated by $s$.
\begin{definition}
Let an extensive game $\mathrm{\Gamma} = \left(N, H, P,
\{\mathcal{I}_{i}\}, \{u_{i}\}\right)$ be given. The normal
representation of $\mathrm{\Gamma}$ is a~strategic game $\left( N,
\{S_{i}\}, \{u_{i}'\} \right)$ in which for each player $i \in N$:
\begin{itemize}
\item $S_{i}$ is the set of pure strategies of a player $i$ in
$\mathrm{\Gamma}$; \item $u_{i}'\colon \prod_{i \in N}S_{i} \to
\mathds{R}$ defined as $u_{i}'(s)\mathrel{\mathop:}=u_{i}(O(s))$
for every $s \in \prod_{i \in N}S_{i}$ and $i \in N$.
\end{itemize}
\end{definition}
One of the most important notions in game theory is a notion of an
equilibrium introduced by John Nash in \cite{nash}. A Nash
equilibrium is a profile of strategies where the strategy of each
player is optimal if the choice of its opponents is fixed. In
other words, in the equilibrium none of the players has any reason
to unilaterally deviate from an equilibrium strategy. A precise
formulation is as follows:
\begin{definition}\label{nashequilibrium}
Let $(N, S_{i}, \{u_{i}\}_{i \in N})$ be a strategic game. A
strategy profile $(t^*_{1}, t^*_{2},\dots,t^*_{n})$ is a Nash
equilibrium (NE) if for each player $i \in N$ and for all $s_{i}
\in S_{i}$:
\begin{align}
u_{i}(t^*_{i}, t^*_{-i}) \geq u_{i}(s_{i}, t^*_{-i}) ~~
\mbox{where} ~~  t^*_{-i} =
(t^*_{1},\dots,t^*_{i-1},t^*_{i+1},\dots,t^*_{n}).\label{nashequation}
\end{align}
\end{definition}
A Nash equilibrium in an extensive game with perfect recall is a
Nash equilibrium of its normal representation, hence
Def.~\ref{nashequilibrium} applies to strategic games as well as
extensive ones. \section{The ultimatum
game}\label{section:ultimatum} 
\begin{figure}[htbp]
\centering
\includegraphics[angle=0, scale=0.8]{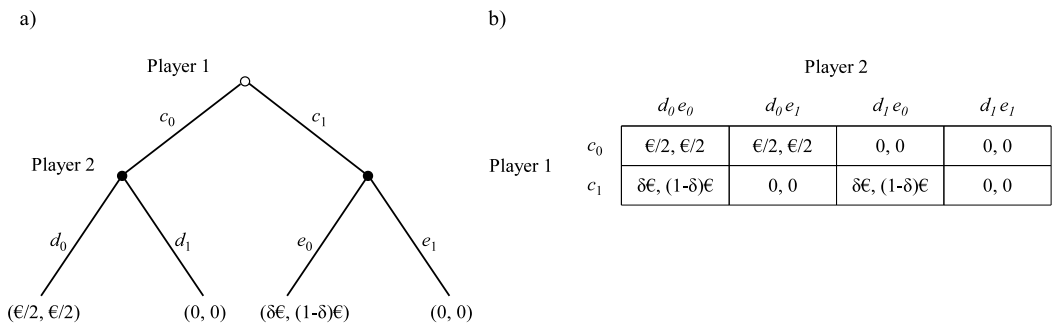}
\caption{A two proposal ultimatum game $\Gamma_{1}$: an extensive
form a) and a~normal form~b).} \label{figure3}
\end{figure}

The ultimatum game is a problem in
which two players face a division of some amount $\geneuro$ of
money. The first player makes the second one a proposal of how to
divide $\geneuro$ between them. Then the second player has to
decide either accept or reject that proposal. The acceptance means
each player receives a part of $\geneuro$ according to the first
player's proposal. If the second player rejects, each player
receives nothing. Let us consider the a variant of the ultimatum
game in which player 1 has two proposals to share $\geneuro$: a
fair division $u_{\mathrm{f}}=(\geneuro/2, \geneuro/2)$ and unfair
one $u_{\mathrm{u}} = (\delta\geneuro, (1-\delta)\geneuro)$, where
the $\delta$ is a fixed factor such that $1/2 < \delta < 1$. This
problem is an extensive game with perfect information that takes
the form:
\begin{align}\label{Gamma1ultimatum}
\Gamma_{1} = \left(\{1,2\}, H, P, \{\mathcal{I}_{i}\}, u\right)
\end{align}
with components defined as follows:
\begin{itemize}
\item $H = \{\emptyset, c_{0}, c_{1}, (c_{0}, d_{0}), (c_{0},
d_{1}), (c_{1}, e_{0}), (c_{1}, e_{1})\}$; \item $P(\emptyset) =
1$,~ $P(c_{0}) = P(c_{1}) = 2$; \item $\mathcal{I}_{1} =
\{\emptyset\}$, $\mathcal{I}_{2} = \{\{(c_{0})\},\{(c_{1})\}\}$;
\item $u(c_{0}, d_{0}) = (\geneuro/2, \geneuro/2)$, ~$u(c_{1},
e_{0}) = (\delta\geneuro, (1-\delta)\geneuro)$,\\ ~$u(c_{0},
d_{1}) = u(c_{1}, e_{1}) = (0,0)$.
\end{itemize}
The extensive and the normal representation of $\Gamma_{1}$ is
shown in Figure~\ref{figure3}. Equilibrium
analysis of the normal representation gives us three pure Nash
equilibria: $(c_{0}, d_{0}e_{1})$, $(c_{1}, d_{0}e_{0})$ and
$(c_{1}, d_{1}e_{0})$. There are also mixed equilibria: a profile
where player 1 chooses $c_{0}$ and player 2 chooses $d_{0}e_{0}$
with probability $p \leq 1/(2\delta)$ and $d_{0}e_{1}$ with
probability $1-p$, and a profile where player 1 decides to play
$c_{1}$ and player 2 chooses any probability distribution over
strategies $d_{0}e_{0}$ and $d_{1}e_{0}$. However, we can put
these ones aside since both mixed equilibria do not contribute to
the utility outcomes of $\Gamma_{1}$. They generate the same
utility outcomes as the pure ones: $(\geneuro/2, \geneuro/2)$ and
$(\delta\geneuro, (1-\delta)\geneuro)$, respectively. The key
feature that make the game $\Gamma_{1}$ so curious is that only
equilibrium profile $(c_{1}, d_{0}e_{0})$ with unfair outcomes
$(\delta\geneuro, (1-\delta)\geneuro)$ is a~reasonable scenario
among all the equilibria of the ultimatum game (many experiments
show that people are inclined to choose fair division
$(\geneuro/2, \geneuro/2)$, however we stick to the natural
assumption of game theory that players are striving to maximize
their payoffs). The strategy combination $(c_{1}, d_{0}e_{0})$ is
the unique equilibrium that is {\em subgame perfect}  (the idea of
subgame perfection is the well-known equilibrium refinement
formulated by Selten \cite{selten}) i.e. it is a profile of
strategies that induces a Nash equilibrium in every subgame (there
are three subgames in $\Gamma_{1}$: the entire game, a game after
the action $c_{0}$ and a~game after the action $c_{1}$). At the
same time the subgame perfection rejects equilibria that are not
credible. Let us consider the profile $(c_{0}, d_{0}e_{1})$. Here,
the strategy $d_{0}e_{1}$ of player 2 demands the action $e_{1}$
when player 1 chooses $c_{1}$. However, when $c_{1}$ occurs, a
rational move of player 2 is $e_{0}$. Similar analysis shows that
also $(c_{1}, d_{1}e_{0})$ is not subgame perfect equilibrium.
Although the notion of subgame perfection is related to the
extensive form of a~game, we can easily determine subgame perfect
equilibria in any two stage extensive game with perfect
information (or even in a wider class of extensive games) through
an analysis of its normal representation. In the game $\Gamma_{1}$
an action taken by player 1 determines a subgame in which only
player 2 makes a move. Thus subgame perfect equilibrium in
$\Gamma_{1}$ is a Nash equilibrium with a property that a strategy
of player 2 is the best response to every strategy of player 1
(i.e., a strategy that weakly dominates the others).
\section{Criticism of the previous approach to \\ quantum
ultimatum game} A misrepresentation of the classical ultimatum
game is the source of its incorrect quantum representation in
\cite{mendes}. The author describes the ultimatum problem as a
$2\times2$ game and then applies the MW and the EWL schemes to
construct the quantum game. However, as we have seen in
Figure~\ref{figure3}b, $2\times4$ is a minimal dimension allowing
to represent the ultimatum game in normal form. A hypothetical
case of the ultimatum game in which player 2 has only two
strategies after an action taken by player 1 implies that player 2
is deprived of capability to make her move conditioned on the
action of the first player. That is tantamount to an event where
the players take their actions at the same time or one of the
players chooses her action as the second but she does not have any
information about an action taken by her opponent. It does not
correspond to a description of the ultimatum game where the second
player knows a proposal of her opponent and depending on the move
of the first player she makes her action. Although the player 2
has only two actions: {\em accept} or {\em reject} in the
two-proposal ultimatum game, in fact she has four pure strategies
defined as her plans of an action at each of her information sets.
Therefore, a $2\times2$ strategic game cannot depict the ultimatum
game. Consequently, the MW and the EWL approach used for
quantization of a $2 \times 2$ game cannot produce a quantum
version of this game. Neither of these quantum realizations
contains the classical ultimatum game.
\section{The quantum ultimatum game obtained by quantization of the normal representation of the classical game}
First, let us remind the protocol for playing quantum games
defined in \cite{fracor3}. It is a six-tuple:
\begin{align}
\Gamma^{\mathrm{QI}} = \left(\mathscr{H}, N,
|\psi_{\mathrm{in}}\rangle, \xi, \{\mathcal{U}_{j}\},
\{E_{i}\}\right) \label{sixtuple}
\end{align}
where the components are defined as follows:
\begin{itemize}
\item $\mathscr{H}$ is a complex Hilbert space $\bigotimes_{j=1}^m
\mathds{C}^2$ with an orthonormal basis $\mathcal{B}$. \item $N$
is a set of players with the property that $|N| \leq m$. \item
$|\psi_{\mathrm{in}}\rangle$ is the initial state of a system of
$m$ qubits $|\varphi_{1}\rangle,
|\varphi_{2}\rangle,\dots,|\varphi_{m}\rangle$. \item $\xi\colon
\{1,2,\dots,m\} \to N$ is a surjective mapping. A value $\xi(j)$
indicates a player who carries out a unitary operation on a qubit
$|\varphi_{j}\rangle$. \item For each $j \in \{1,2,\dots,m\}$ the
set $\mathcal{U}_{j}$ is a subset of unitary operators from
$\mathsf{SU}(2)$ that are available for a qubit $j$. A~(pure)
strategy of a player $i$ is a map $\tau_{i}$ that assigns a
unitary operation $U_{j} \in \mathcal{U}_{j}$ to a qubit
$|\varphi_{j}\rangle$ for every $j \in \xi^{-1}(i)$. The final
state $|\psi_{\mathrm{fin}}\rangle$ when the players have
performed their strategies on corresponding qubits is defined as:
\begin{align}
|\psi_{\mathrm{fin}}\rangle\mathrel{\mathop:}=
(\tau_{1},\tau_{2},\dots,\tau_{n})|\psi_{\mathrm{in}}\rangle =
\bigotimes_{i\in N}\bigotimes_{j\in \xi^{-1}(i)}U_{j}
|\psi_{\mathrm{in}}\rangle. \label{finalstate}
\end{align} \item For each $i\in N$ the map $E_{i}$ is a utility (payoff) functional that
specifies a utility for the player $i$. The functional $E_{i}$ is
defined by the formula:
\begin{align}\label{eformula}
E_{i} = \sum_{|b\rangle \in \mathcal{B}}v_i(b)|\langle b
|\psi_{\mathrm{fin}}\rangle|^2, ~~ \mbox{where} ~~ v_i(b) \in
\mathds{R}.
\end{align}
\end{itemize}
The above scheme is adapted for extensive games with two available
actions at each information set so that we could use only qubits
for convenience. Any game richer in actions can be transferred to
quantum domain by using quantum objects of higher dimensionality.

The idea framed in \cite{fracor3} bases on identifying unitary
actions taken on a qubit with actions taken in an information set
of classical game. Therefore, three qubits are required to express
the ultimatum game in quantum information language. Since the
first player has one information set and the second player has two
ones, player 1 performs a unitary operation on only one qubit and
player 2 operates on the rest. Like in \cite{mendes} we examine
the two approaches: the MW approach and the EWL approach to
quantizing $\Gamma_{1}$.
\subsection{The MW approach}\label{subsection_MW} 
\begin{figure}[t]
\includegraphics[angle=0, scale=0.8]{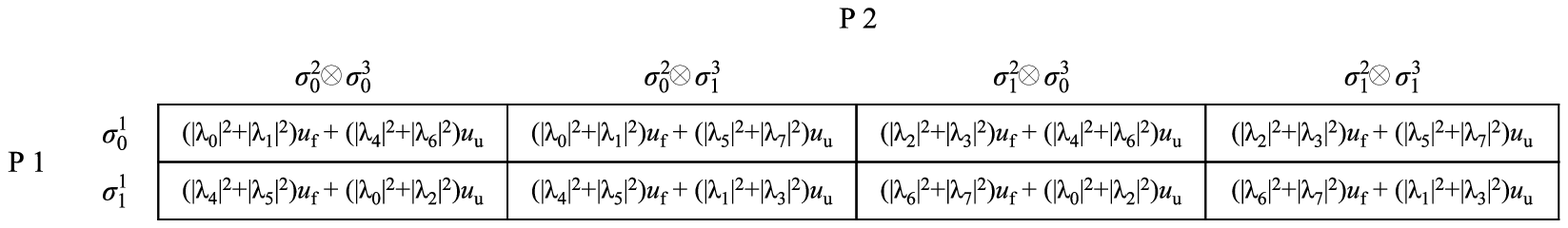}
\caption{The MW approach to the normal representation of
$\Gamma_{1}$.} \label{figure4}
\end{figure}

Let us consider the following
six-tuple:
\begin{align}
\Gamma^{\mathrm{MW}}_{1} = \left(\mathscr{H}_{c}, \{1,2\},
|\psi_{\mathrm{in}}\rangle, \xi, \{\{\sigma_{0},
\sigma_{1}\}_{i}\}, \{E_{i}\} \right), \label{mw2}
\end{align}
where:
\begin{itemize}
\item  $\mathscr{H}_{c}$ is a Hilbert space
$\bigotimes^3_{j=1}\mathds{C}^2$ with the computational basis
states $|x_{1},x_{2},x_{3}\rangle$, $x_{j}=0,1$; \item the initial
state $|\psi_{\mathrm{in}}\rangle$ is a general pure state of
three qubits:
\begin{align}
|\psi_{\mathrm{in}}\rangle = \sum_{x \in \{0,1\}^3} \lambda_{x}|x
\rangle, ~~\mbox{where}~~ \lambda_{x} \in \mathds{C}
~~\mbox{and}~~ \sum_{x \in \{0,1\}^3}|\lambda_{x}|^2 = 1;
\label{generalinitialstate}
\end{align}
\item the map $\xi$ on $\{1,2,3\}$ given by the formula: $\xi(j) =
\left\{\begin{array}{lll}
1, & \mbox{if} & j=1;\\
2, & \mbox{if} & j \in \{2,3\}.
\end{array} \right.$
\item $\sigma_{0} = \left(\begin{array}{cc} 1 &
 0 \\
 0  & 1
\end{array}\right)$ \,and\, $\sigma_{1} = \left(\begin{array}{cc} 0 &
 1 \\
 1  & 0
\end{array}\right)$;
  \item the payoffs functionals $E_{i}$, $i=1,2$, are of
the form:
\begin{align} \begin{split}
&E_{1} =
\frac{1}{2}\geneuro\sum_{x_{3}}|\langle00,x_{3}|\psi_{\mathrm{fin}}\rangle|^2
+ \delta\geneuro\sum_{x_{2}}|\langle
1,x_{2},0|\psi_{\mathrm{fin}}\rangle|^2;\\ &E_{2} =
\frac{1}{2}\geneuro
\sum_{x_{3}}|\langle00,x_{3}|\psi_{\mathrm{fin}}\rangle|^2 +
(1-\delta)\geneuro\sum_{x_{2}}|\langle
1,x_{2},0|\psi_{\mathrm{fin}}\rangle|^2. \end{split}
\label{xoperators}
\end{align}
\end{itemize}
By definition of $\xi$ in $\Gamma_{1}$, player 1 acts on the first
qubit and treats the operators $\sigma^1_{0}$ and $\sigma^1_{1}$
as her strategies. Player 2 acts on the second and the third
qubit, hence her pure strategies are $\sigma^2_{0} \otimes
\sigma^3_{0}$, $\sigma^2_{0} \otimes \sigma^3_{1}$, $\sigma^2_{1}
\otimes \sigma^3_{0}$ and $\sigma^2_{1} \otimes \sigma^3_{1}$ (the
upper index denotes a qubit on which an operation is made). Let us
determine for each profile $\left(\sigma^1_{\kappa_{1}},
\left(\sigma^2_{\kappa_{2}}, \sigma^3_{\kappa_{3}}\right)\right)$,
where $\kappa_{1}, \kappa_{2}, \kappa_{3} \in \{0,1\}$, the
corresponding expected utility $E_{i}$ by using formulae
(\ref{finalstate})-(\ref{eformula}) and the specification of
(\ref{mw2}). We illustrate it using as an example
$E_{i}\left(\sigma^1_{0}, \left(\sigma^2_{1},
\sigma^3_{0}\right)\right)$ for $i=1,2$. The initial state after
the players choose the profile
$\left(\sigma^1_{0},\left(\sigma^2_{1},
\sigma^3_{0}\right)\right)$ takes the form
$|\psi_{\mathrm{fin}}\rangle=\sigma^1_{0}\otimes\sigma^2_{1}\otimes\sigma^3_{0}|\psi_{\mathrm{in}}\rangle.$
Thus, we have:
\begin{align} \label{finalstatejeden}
|\psi_{\mathrm{fin}}\rangle = \sum_{x_{1},x_{2},x_{3} \in \{0,1\}}
\lambda_{x_{1},x_{2},x_{3}}|x_{1},\overline{x}_{2}, x_{3}\rangle,
\end{align}
where $\overline{x}_{2}$ is the negation of $x_{2}$. Putting the
final state (\ref{finalstatejeden}) into the first of
Eq.~(\ref{xoperators}) we obtain:
\begin{align}
E_{1}\left(\sigma^1_{0},\left(\sigma^2_{1},
\sigma^3_{0}\right)\right) =
\frac{1}{2}\geneuro\left(|\lambda_{010}|^2 +
|\lambda_{011}|^2\right) +
\delta\geneuro\left(|\lambda_{100}|^2+|\lambda_{110}|^2 \right).
\end{align}
Obviously, we have $(1-\delta)\geneuro$ instead of
$\delta\geneuro$ in the expected utility $E_{2}$. Therefore, the
payoff vector $(E_{1}, E_{2})$ is
$u_{\mathrm{f}}\left(|\lambda_{010}|^2 + |\lambda_{011}|^2\right)
+ u_{\mathrm{u}}\left(|\lambda_{100}|^2 +
|\lambda_{110}|^2\right)$ in that case. Payoff vectors $(E_{1},
E_{2})$ for all possible profiles $\left(\sigma^1_{\kappa_{1}},
\left(\sigma^2_{\kappa_{2}}, \sigma^3_{\kappa_{3}}\right)\right)$
are placed in the matrix representation in Figure~\ref{figure4}
(for convenience we convert binary indices $(x_{1}, x_{2},
x_{3})_{2}$ of $\lambda_{x_{1},x_{2}, x_{3}}$ to the decimal
numeral system).

Let us examine the game in Figure~\ref{figure4} to answer to what
degree passing to the quantum domain may influence the result of
the game. Notice first that (\ref{mw2}) is indeed the quantum game
in the spirit of the MW approach - the normal representation of
$\Gamma_{1}$ can be obtained from $\Gamma^{\mathrm{MW}}_{1}$ by
putting $|\lambda_{0}|^2 = 1$ and $|\lambda_{x}|^2 = 0$ for
$x=1,2,\dots ,7$, i.e., if we put $|\psi_{\mathrm{in}}\rangle =
|000\rangle$. More generally: $\Gamma^{\mathrm{MW}}_{1}$ coincides
to a game isomorphic to the normal representation of $\Gamma_{1}$
if we put as $|\psi_{\mathrm{in}}\rangle =
|x_{1},x_{2},x_{3}\rangle$ any basis state. Then
$\Gamma^{\mathrm{MW}}_{1}$ is equal to $\Gamma_{1}$ up to the
order of players' strategies. The game $\Gamma_{1}$ favors player
1 as we have learnt in Section~\ref{section:ultimatum} Thus, an
interesting problem is to look for another form of the initial
state (\ref{generalinitialstate}) that imply fairer solution
unavailable in the game $\Gamma_{1}$. Let us study first:
\begin{align}
|\psi_{\mathrm{in1}}\rangle = \frac{1}{2}\left(|000\rangle +
|001\rangle + |100\rangle + |110\rangle \right).
\label{initialstate1}
\end{align}
Through the substitution $|\lambda_{0}|^2 = |\lambda_{1}|^2 =
|\lambda_{4}|^2 = |\lambda_{6}|^2 = 1/4$ (the other squares of the
moduli equal 0) to entries of the matrix representation in
Figure~\ref{figure4} we obtain a game where the only reasonable
equilibrium profile is $\sigma^1_{0} \otimes \sigma^2_{0} \otimes
\sigma^3_{0}$ with corresponding expected utility vector $E =
(E_1, E_{2})$ equal $\left(u_{\mathrm{f}} +
u_{\mathrm{u}}\right)/2$. The other pure equilibria: $\sigma^1_{1}
\otimes \sigma^2_{0} \otimes \sigma^3_{1}$ and $\sigma^1_{1}
\otimes \sigma^2_{1} \otimes \sigma^3_{1}$ - both generating the
utility outcome $\left(u_{\mathrm{f}} + u_{\mathrm{u}}\right)/4$
are obviously worse for both players so they won't be chosen.
Moreover, $\sigma^1_{0} \otimes \sigma^2_{0} \otimes \sigma^3_{0}$
is an imitation of a subgame perfect equilibrium - the strategy of
the second player $\sigma^2_{0}\otimes\sigma^3_{0}$ is the best
response to any strategy of the first player. To sum up, the
initial state (\ref{initialstate1}) is beneficial to player 2
compared with the classical case.
 It turns out that the answer to the question: is
there any $|\psi_{\mathrm{in}}\rangle$ allowing to obtain a fair
division of $\geneuro$, is also positive. Let us consider any
state of the form:
\begin{align}
|\psi_{\mathrm{in2}}\rangle = \sqrt{\frac{1}{2\delta'}}|000\rangle
+ \sqrt{1-\frac{1}{2\delta'}}|001\rangle, ~~ \mbox{where}
~~\frac{1}{2} < \delta < \delta' < 1. \label{initialstate2}
\end{align}
Once again the profile $\sigma^1_{0} \otimes \sigma^2_{0} \otimes
\sigma^3_{0}$ constitutes a Nash equilibrium and the strategy of
the second player $\sigma^2_{0} \otimes \sigma^3_{0}$ weakly
dominates her other strategies as a result of putting
$|\lambda_{0}|^2 = 1/2\delta'$ and $|\lambda_{1}|^2 =
1-1/2\delta'$ in the game in Figure~\ref{figure4}. Since there are
no other profiles with that property, $\sigma^1_{0} \otimes
\sigma^2_{0} \otimes \sigma^3_{0}$ is the most reasonable scenario
that implies $E_{1,2}(\sigma^1_{0} \otimes \sigma^2_{0} \otimes
\sigma^3_{0}) =\geneuro/2$. The superposition of the third qubit
(the second qubit of player 2) is essential to obtain fair
division result since it is impossible to achieve $\delta\geneuro$
by player 1 then. Therefore, the payoff $\geneuro/2$ becomes the
most attractive for her now.

The conclusions we can draw from the analysis of the MW approach
to the ultimatum game are as follows. First, the game
$\Gamma^{\mathrm{MW}}_{1}$ that begins with
$|\psi_{\mathrm{in}1}\rangle$ discloses a game tree different from
the one in Figure~\ref{figure3}a). If there is a protocol for
quantizing the extensive game $\Gamma_{1}$ directly without using
its normal representation as in our case, then the output game
tree must be different from the game tree of $\Gamma_{1}$ in
general. It follows form the fact that the game tree in
Figure~\ref{figure3}a with any four utility outcomes assigned to
its terminal histories implies the normal representation specified
by only these four payoff outcomes. However, the game
$\Gamma^{\mathrm{MW}}_{1}$ where the initial state take the form
of (\ref{initialstate1}) has five different outcomes. Notice, that
is not irrelevant issue bearing in mind the fact that the bimatrix
of a strategic game played classically as well as played by the MW
protocol always have the same dimension.

The case where game begins with the state (\ref{initialstate2}) is
applied shows that even a separable initial state can influence
significantly a result of $\Gamma_{1}$. It is not strange
property. Any superposition of a player's qubit causes some
limitation on players' influence on their qubits as we have seen
in the case (\ref{initialstate1}). In particular, if each qubit of
the initial state is in the state $|+\rangle = \left(|0\rangle +
|1\rangle\right)/\sqrt{2}$, no player can affect amplitudes of her
qubit applying only $\sigma_{0}$ and $\sigma_{1}$ (measurement
outcomes 0 and 1 on qubit occur with the same probability). Then
the result of the game only depends on the initial state
$|\psi_{\mathrm{in}}\rangle = |+\rangle|+\rangle|+\rangle$.
\subsection{The EWL approach} As we have seen, the two-element set
of unitary operators is too simple in some cases. The
two-parameter unitary operations used in the EWL protocol allow to
avoid player's powerlessness when she acts on $|+\rangle$, and
generally each player can essentially affect amplitudes of the
initial state. Thus, it is interesting to find a result of the
ultimatum game played according to the EWL approach. Let the
following six-tuple be given:
\begin{align}\label{gammaewl}
\Gamma^{\mathrm{EWL}}_{1} = \left(\mathscr{H}_{e}, \{1,2\},
|\psi_{000}\rangle, \xi, \{\{U(\theta,\beta)\}_{i}\},
\{E_{i}\}\right),
\end{align}
where:
\begin{itemize}
\item $\mathscr{H}_{e}$ is a Hilbert space $\bigotimes_{j=1}^3
\mathds{C}^2$ with the basis $\{|\psi_{x_{1}, x_{2},
x_{3}}\rangle\colon x_{j} = 0,1\}$ of entangled states defined as
follows:
\begin{eqnarray}
|\psi_{x_{1},x_{2},x_{3}}\rangle = \frac{|x_{1},x_{2},x_{3}\rangle
+
i|\overline{x}_{1},\overline{x}_{2},\overline{x}_{3}\rangle}{\sqrt{2}};
\label{basestate}
\end{eqnarray} \item the mapping $\xi$ is the same as in six-tuple (\ref{mw2}); \item
the player's actions $\{U(\theta,\beta)\colon \theta \in [0,\pi],
\beta \in [0,\pi/2]\}$, studied, for example, in the paper
\cite{flitney}, form an alternative to two-parameter unitary
operations used in \cite{eisert}. They are of the form:
\begin{align}
U(\theta, \beta) = \left(\begin{array}{cc} \cos(\theta/2) &
 ie^{i\beta}\sin(\theta/2) \\
 ie^{-i\beta}\sin(\theta/2)  & \cos(\theta/2)
\end{array}\right); \label{twoparameter}
\end{align}
\item $E_{i}$ for $i=1,2$ are the payoff functionals
(\ref{xoperators}) defined for the basis (\ref{basestate}):
\begin{align}
\begin{split}
&E_{1} = \frac{1}{2}\geneuro \sum_{x_{3}} |\langle
\psi_{00{,}x_{3}}|\psi_{\mathrm{fin}}\rangle|^2 + \delta\geneuro
\sum_{x_{2}} |\langle\psi_{1{,}x_{2}{,}0}|\psi_{\mathrm{fin}}\rangle|^2;\\
&E_{2} = \frac{1}{2}\geneuro \sum_{x_{3}} |\langle
\psi_{00{,}x_{3}}|\psi_{\mathrm{fin}}\rangle|^2 +
(1-\delta)\geneuro \sum_{x_{2}}
|\langle\psi_{1{,}x_{2}{,}0}|\psi_{\mathrm{fin}}\rangle|^2.
\label{x2operators}
\end{split}
\end{align}
\end{itemize}
Each strategy $U_{1}$ of player 1 is simply $U(\theta_{1},
\beta_{1})$.  The strategies of the second player are chosen in a
manner similar to $\Gamma^{\mathrm{MW}}_{1}$ - they are tensor
products $U_{2} \otimes U_{3} = U(\theta_{2}, \beta_{2}) \otimes
U(\theta_{3}, \beta_{3})$. The final state
$|\psi_{\mathrm{fin}}\rangle$ corresponding to a~profile $\tau =
((\theta_{1}, \beta_{1}), (\theta_{2}, \beta_{2}, \theta_{3},
\beta_{3}))$ is as follows:
\begin{align}
|\psi_{\mathrm{fin}}\rangle = U_{1}\otimes U_{2} \otimes
U_{3}|\psi_{000}\rangle = \frac{1}{\sqrt{2}}\sum_{x \in
\{0,1\}^3}\upsilon_{x}|x\rangle, \label{rhogamma1}
\end{align}
where
\begin{align}
\upsilon_{x_{1},x_{2},x_{3}} &= i^{\sum x_{j}}e^{-i\sum
x_{j}\beta_{j}}\prod_{j}\cos\left(\frac{x_{j}\pi -
\theta_{j}}{2}\right)\notag\\&\quad+(-i)^{\sum x_{j}}e^{i\sum
\overline{x_{j}}\beta_{j}}\prod_{j}\cos\left(\frac{\overline{x_{j}}\pi
- \theta_{j}}{2}\right),
\end{align}
and $j=1,2,3$, $x_{j} = 0,1$, and $\overline{x}_{j}$ is negation
of $x_{j}$. Putting (\ref{x2operators}) and (\ref{rhogamma1}) into
formula (\ref{eformula}) we obtain the following expected payoff
vector $(E_{1}(\tau), E_{2}(\tau))$:
\begin{align}\label{payoffewl}
&(E_{1}(\tau), E_{2}(\tau))= \notag\\
&\qquad
u_{\mathrm{f}}\Biggl[\cos^2\frac{\theta_{1}}{2}\cos^2\frac{\theta_{2}}{2}\left(\cos^2\frac{\theta_{3}}{2}
+ \sin^2\frac{\theta_{3}}{2}\cos^2\beta_{3}\right)\notag\\
&\qquad+\sin^2\frac{\theta_{1}}{2}\sin^2\frac{\theta_{2}}{2}\left(\sin^2\frac{\theta_{3}}{2}\sin^2(\beta_{1}
+ \beta_{2} + \beta_{3}) +
\cos^2\frac{\theta_{3}}{2}\sin^2(\beta_{1} + \beta_{2})
\right)\Biggr]\notag\\
&\qquad+u_{\mathrm{u}}\Biggl[\sin^2\frac{\theta_{1}}{2}\cos^2\frac{\theta_{3}}{2}\left(\cos^2\frac{\theta_{2}}{2}\cos^2\beta_{1}
+ \sin^2\frac{\theta_{2}}{2}\cos^2(\beta_{1} + \beta_{2})\right)\notag\\
&\qquad+\cos^2\frac{\theta_{1}}{2}\sin^2\frac{\theta_{3}}{2}\left(\sin^2\frac{\theta_{2}}{2}\sin^2(\beta_{2}
+ \beta_{3}) +
\cos^2\frac{\theta_{2}}{2}\sin^2\beta_{3}\right)\Biggr].
\end{align}
Let us check first that $\Gamma^{\mathrm{EWL}}_{1}$ generalizes
the classical ultimatum game $\Gamma_{1}$. Pure strategies of the
first player are represented by $U(0,0)$ and $U(\pi,0)$.
Similarly, the set of strategies of the second player in
$\Gamma_{1}$ is represented by a set $\{U(\theta_{2},0)\otimes
U(\theta_{3},0)\colon \theta_{2}, \theta_{3} \in \{0,\pi\}\}$
since the set of profiles
\begin{align}
\{\left((\theta_{1},0),(\theta_{2},0,\theta_{3},0)\right)\colon
\theta_{1},\theta_{2},\theta_{3}\in \{0,\pi\}\} \end{align} in
(\ref{gammaewl}) and the set of profiles
\begin{align}
\{(c_{k_{1}},d_{k_{2}}e_{k_{3}}) \colon k_{1},k_{2},k_{3} \in
\{0,1\}\} \end{align} in (\ref{Gamma1ultimatum}) generate the same
payoffs. Equivalents of behavioral strategies of $\Gamma_{1}$
(i.e., independent probability distributions $p$, $q$ and $r$ over
the actions $c_{k_{1}}$, $d_{k_{2}}$ and $e_{k_{3}}$,
respectively, specified by players at their own information sets)
can be found among unitary strategies as well. If we restrict
unitary actions to $U(\theta, 0)$, i.e., to profiles of the form
 $((\theta_{1},0), (\theta_{2},0,\theta_{3},0))$, $\theta_{j}
\in [0,\pi]$, the right-hand side of Eq.~(\ref{payoffewl}) takes
the form:
\begin{align}
u_{\mathrm{f}}\cos^2\frac{\theta_{1}}{2}\cos^2\frac{\theta_{2}}{2}
+
u_{\mathrm{u}}\sin^2\frac{\theta_{1}}{2}\cos^2\frac{\theta_{3}}{2}.
\end{align}
By substituting $p$ for $\cos^2(\theta_{1}/2)$, $q$ for
$\cos^2(\theta_{2}/2)$, and $r$ for $\cos^2(\theta_{3}/2)$ we get
the expected payoffs corresponding to any behavioral strategy
profile $((p,1-p),((q,1-q), (r,1-r)))$ in $\Gamma_{1}$.

Let us examine an impact of the unitary strategies on a result of
the EWL approach to $\Gamma_{1}$.  In particular we ask the
question if the unfair division $u_{\mathrm{u}}$ or the fair
division $u_{\mathrm{f}}$ in $\Gamma^{\mathrm{EWL}}_{1}$ is more
probable. Notice, that the profile $((\theta_{1}, \beta_{1}),
(\theta_{2}, \beta_{2}, \theta_{3},\beta_{3})) = ((\pi, 0), (0, 0,
0, 0))$ (corresponding to subgame perfect equilibrium $(c_{1},
d_{0}e_{0})$ in $\Gamma_{1}$) is not Nash equilibrium in
$\Gamma^{\mathrm{EWL}}_{1}$. The second player can gain by
choosing, for example, $(\theta_{2}, \beta_{2},
\theta_{3},\beta_{3}) = (\pi,\pi/2, \pi, 0)$ instead of $(0, 0, 0,
0)$. Then she obtains the fair devision payoff. Moreover, for any
other strategy of the first player $(\theta_{1}, \beta_{1})$,
player 2 can select, for instance, $(0, 0, 0, 0)$ to obtain a
payoff being a mixture of $u_{\mathrm{f}}$ and $u_{\mathrm{u}}$.
This proves that the unfair division $u_{\mathrm{u}}$ cannot be a
result in (\ref{gammaewl}). The fair division $u_{\mathrm{f}}$ in
turn can be achieved through continuum of Nash  equilibria. Let us
denote by $\mathrm{NE}(\Gamma^{\mathrm{EWL}}_{1})$ the set of all
Nash equilibria of $\Gamma^{\mathrm{EWL}}_{1}$. An examination of
(\ref{payoffewl}) shows that:
\begin{align}
\left\{((\pi,\beta_{1}),(\pi,\beta_{2},\pi, \beta_{3})) \colon
\beta_{2}+ \beta_{3} \leq \frac{\pi}{4}\,, \sum^3_{j = 1}\beta_{j}
= \frac{\pi}{2}\right\} \subset
\mathrm{NE}(\Gamma^{\mathrm{EWL}}_{1}) \label{subset1}
\end{align}
as well as
\begin{align}
\left\{((0,\beta_{1}),(0,\beta_{2},\pi,0)) \colon \beta_{1},
\beta_{2} \in \left[0,\frac{\pi}{2}\right]\right\} \subset
\mathrm{NE}(\Gamma^{\mathrm{EWL}}_{1}). \label{subset2}
\end{align}
Moreover, all strategy profiles of these sets generate the payoff
vector $u_{\mathrm{f}}$ for any division factor $1/2<\delta<1$. To
prove inclusion (\ref{subset1}) let us consider any strategy
$(\theta'_{1},\beta'_{1})$ of player 1 given that player 2's
strategy from (\ref{subset1}) is fixed. Then for
$\beta_{2}+\beta_{3}\leq \pi/4$ we have
\begin{multline}\label{multiequation}
E_{1}((\theta'_{1},\beta'_{1}),(\pi,\beta_{2},\pi,\beta_{3}))\\
=
\left[\frac{1}{2}\geneuro\sin^2\frac{\theta'_{1}}{2}\sin^2(\beta'_{1}+\beta_{2}+\beta_{3})+
\delta\geneuro\cos^2\frac{\theta'_{1}}{2}\sin^2(\beta_{2}+\beta_{3})\right].
\end{multline}
Since $\beta_{2}+\beta_{3}\leq \pi/4$, the maximum value of
(\ref{multiequation}) is achieved if the second element of the sum
is 0. It implies that the best response of player 1 is
$\theta'_{1} = \pi$ and $\beta_{1}'= \pi/2 - \beta_{2}-\beta_{3}$.
The second player cannot gain by deviating as well because she
always obtains no more than $\geneuro/2$ in
$\Gamma^{\mathrm{EWL}}_{1}$. Therefore, each profile of set
(\ref{subset1}) indeed constitutes Nash equilibrium. Inclusion
(\ref{subset2}) can be proved in similar way. Notice that there
are also Nash equilibria different from (\ref{subset1}) and
(\ref{subset2}) that generate the payoff outcome $\geneuro/2$ for
both players. For example, a strategy profile $((\pi, \pi/4),(\pi,
\pi/4, \pi/2,0))$.

Intuitively, a huge number of fair solutions in
$\Gamma^{\mathrm{EWL}}_{1}$ being NE together with a lack of an
equilibrium outcome $u_{\mathrm{u}}$ favors the second player in
comparison to the classical game $\Gamma_{1}$ . However, it does
not assure the second player the fair payoff $\geneuro/2$ yet.
Since the players choose their strategies simultaneously, they
cannot coordinate them. If the first player unilaterally deviates
from a strategy dictated by (\ref{subset2}) and she plays a
strategy being a part of (\ref{subset1}) then both players receive
nothing as we have $E_{1,2}((\pi,\beta_{1}), (0, \beta_{2}, \pi,
0)) = 0$ for all $\beta_{1}, \beta_{2} \in \left[0,\pi/2\right]$.
On the other hand, it turns out that the statement that each of
these equilibria is equally likely to occur is not true. Let us
investigate which equilibria in $\Gamma_{1}$ are preserved in
$\Gamma^{\mathrm{EWL}}_{1}$ bearing in mind that the unitary
strategies $U(\theta,0)$ are quantum counterparts to classical
moves in $\Gamma_{1}$. As we have seen there is no equilibrium
profile in $\Gamma^{\mathrm{EWL}}_{1}$ that allows the first
player to gain $\delta\geneuro$. Therefore, in particular, the
unfair division equilibrium $(c_{1},d_{0}e_{0})$ of $\Gamma_{1}$
cannot be generated by any unitary operations $U(\theta,0)$.
However, each fair division equilibrium (pure or mixed) of
(\ref{Gamma1ultimatum}) can be reconstructed in (\ref{gammaewl}).
The profile $((0,0),(0,0,\pi,0))$ corresponding to the equilibrium
$(c_{0},d_{0}e_{1})$ in $\Gamma_{1}$ is Nash equilibrium of
$\Gamma^{\mathrm{EWL}}_{1}$ since it is element of the set
(\ref{subset2}). Next, the mixed equilibria mentioned in Section
\ref{section:ultimatum} can be implemented in
$\Gamma^{\mathrm{EWL}}_{1}$ as follows: they are the profiles
where the first player chooses $U(0,0)$ and the second player
chooses either $U(0,0)\otimes U(0,0)$ with probability $p \in [0,
1/2\delta]$ and $U(0,0)\otimes U(\pi,0)$ with probability $1-p$,
or in a language of behavioral strategies she just takes an
operator from $\bigl\{U(0,0) \otimes U(\theta,0)\colon \theta \in
\bigl[2\arccos\bigl(1/\sqrt{2\delta}\bigr), \pi/2\bigr]\bigr\}$.
 According to the
concept of Schelling Point \cite{schelling} players tend to select
a solution that is the most natural as well as the most
distinctive among all possible choices. Therefore, if we assume
that the players prefer the fair division, they choose a profile
that is an equilibrium of both $\Gamma_{1}$ and
$\Gamma^{\mathrm{EWL}}_{1}$ among all equal equilibria of
$\Gamma^{\mathrm{EWL}}_{1}$. Since all these shared equilibria
generate the same outcome, the pure equilibrium is the most
natural and it ought to be chosen as the Schelling Point.
\section{Extensive form of the quantum ultimatum game}
In subsection~\ref{subsection_MW} we made observation that an
extensive game and its quantum realization differ not only in
utilities but also in game trees. Now, we are going to give the
answer to the question how would a game tree of such quantum
realization look like? Let us reconsider an extensive game form
given by the game tree on Figure~\ref{figure3}a, where the
components $H$, $P$ and $\mathcal{I}_{i}$ are derived from
$\Gamma_{1}$, and the outcomes $O_{00}, O_{01}, O_{10}$ and
$O_{11}$, are assigned to the terminal histories $(c_{0}, d_{0}),
(c_{0}, d_{1}), (c_{1}, e_{0})$ and $(c_{1}, e_{1})$,
respectively, instead of particular payoff values. Let us denote
this problem as:
\begin{align}
\Gamma_{2} = \left(\{1,2\}, H, P, \mathcal{I}_{i}, O\right).
\label{gamma3}
\end{align}
Then the tuple $\Gamma^{\mathrm{MW}}_{2}$ associated with
$\Gamma_{2}$ is derived from $\Gamma^{\mathrm{MW}}_{1}$ and only
the payoff functionals $E_{i}$ undergo appropriate modifications.
Let us write $\Gamma^{\mathrm{MW}}_{2}$ in the language of density
matrices, for convenience. That is:
\begin{align}
\Gamma^{\mathrm{MW}}_{2} = \left(\mathscr{H}_{c}, \{1,2\},
\rho_{\mathrm{in}}, \xi, \{\sigma_{0}, \sigma_{1}\}_{i}, X\right),
\label{gameMW3}
\end{align}
where
\begin{itemize}
\item $\rho_{\mathrm{in}}$ is a density matrix of the initial
state (\ref{generalinitialstate}); \item the outcome operator $X$
is a sum of $X^{0} + X^{1}$ defined as:
\begin{align} \begin{split}
&X^0 = O_{00}|00\rangle \langle 00| \otimes \mathds{1} +
O_{01}|01\rangle
\langle 01| \otimes \mathds{1};\\
&X^1 = O_{10}|1\rangle \langle 1| \otimes \mathds{1} \otimes
|0\rangle \langle 0| + O_{11}|1\rangle \langle 1| \otimes
\mathds{1} \otimes |1\rangle \langle 1|. \end{split}
\end{align}
\end{itemize}
In this case, the density matrix $\rho_{\mathrm{fin}}$ of the
final state $|\psi_{\mathrm{fin}}\rangle$ takes a form
\begin{align}\label{finalstateultimatum}
\rho_{\mathrm{fin}} = \sigma^1_{\kappa_{1}} \otimes
\sigma^2_{\kappa_{2}}\otimes \sigma^3_{\kappa_{3}}
\rho_{\mathrm{in}} \sigma^1_{\kappa_{1}} \otimes
\sigma^2_{\kappa_{2}}\otimes \sigma^3_{\kappa_{3}}.
\end{align}
The outcome functionals (\ref{eformula}) are then equivalent to
the following one:
\begin{align} \label{eformulaultimatum}
E\left( \sigma^1_{\kappa_{1}},\left(\sigma^2_{\kappa_{2}},
\sigma^3_{\kappa_{3}}\right)  \right) =
\mathrm{tr}\left(X\rho_{\mathrm{fin}}\right).
\end{align}
In order to give a~extensive form to determine the final state
$\rho_{\mathrm{fin}}$ in $\Gamma^{\mathrm{MW}}_{2}$ let us modify
the way (\ref{finalstateultimatum}) of calculating the final state
$\rho_{\mathrm{fin}}$. To begin with, player 1 acts on the first
qubit. Next, player 2 carries out a measurement on that qubit in
the computational basis to find out what is a current state of the
game. Then she performs an operation on either the second or the
third qubit of the post-measurement state depending on whether the
measurement outcome 0 or 1 has occurred. The operation of the
second player ultimately defines the final state that is inserted
to the formula~(\ref{eformulaultimatum}). The procedure can be
formalized as follows:\\

\newcommand*{\dupaa}{}
\newcommand*{\tempdupa}{\bf Sequential procedure}
\newcommand*{\pustya}{}
\newcommand*{\tempDupa}{}
\newcommand*{\dupaI}{1.}
\newcommand*{\temphead}{$\sigma^1_{\kappa_{1}}\rho_{\mathrm{in}}\sigma^1_{\kappa_{1}} = \rho_{\kappa_{1}}$}
\newcommand*{\pustyI}{}
\newcommand*{\TempHead}{\footnotesize the player 1 performs an operation $\sigma^1_{\kappa_{1}}$ on her
qubit of the initial state $\rho_{\mathrm{in}}$}
\newcommand*{\dupaII}{2.}
\newcommand*{\tempdrugi}{$\displaystyle\frac{M_{\iota}\rho_{\kappa_{1}} M_{\iota}}{\mathrm{tr}(M_{\iota}\rho_{\kappa_{1}})} = \rho_{\kappa_{1},\iota}$, \newline $p_{\kappa_{1},\iota} =
\mathrm{tr}(M_{\iota}\rho_{\kappa_{1},\iota}$)}
\newcommand*{\pustyII}{}
\newcommand*{\Tempdrugi}{\footnotesize the player 2 prepares the measurement $\{M_{0},
M_{1}\}$ defined by $M_{\iota} = |\iota \rangle \langle \iota|
\otimes I \otimes I, \, \iota =0,1$ on the first qubit of the
state $\sigma_{\kappa_{1}}\rho_{\mathrm{in}}\sigma_{\kappa_{1}}$
(the probability of obtaining result $\iota$ is denoted by
$p_{\kappa_{1},\iota}$)}
\newcommand*{\dupaIII}{3.}
\newcommand*{\temptrzeci}{$\sum_{\iota} p_{\kappa_{1},\iota}\sigma^{2+\iota}_{\kappa_{2+\iota}} \rho_{\kappa_{1},\iota} \sigma^{2+\iota}_{\kappa_{2+\iota}} =
\rho'_{\mathrm{fin}}$}
\newcommand*{\pustyIII}{}
\newcommand*{\Temptrzeci}{\footnotesize if a measurement outcome $\iota$ occurs, the player 2 performs an operaton $\sigma_{\kappa_{\iota+2}}$ on $\iota + 2$ qubit of the post-measurement state}
\newcommand*{\dupaIV}{}
\newcommand*{\tempczwarty}{}
\newcommand*{\pustyIV}{}
\newcommand*{\Tempczwarty}{}
\setlength{\tabcolsep}{-5mm}
\renewcommand*{\arraystretch}{1.5}
\begin{tabular}{p{2cm}p{5cm}p{2.5cm}p{8.5cm}}
\dupaa & \tempdupa & \pustya & \tempDupa\\
\dupaI & \temphead & \pustyI & \TempHead\\
\dupaII & \tempdrugi & \pustyII & \Tempdrugi\\
\dupaIII & \temptrzeci & \pustyIII & \Temptrzeci\\
\dupaIV & \tempczwarty & \pustyIV & \Tempczwarty
\end{tabular}

\noindent It turns out that for any strategy profile
$\left(\sigma^1_{\kappa_{1}}, \left(\sigma^2_{\kappa_{2}},
\sigma^3_{\kappa_{3}}\right)\right)$ the final state
$\rho_{\mathrm{fin}}$ defined both by the formula
(\ref{finalstateultimatum}) and by the sequential procedure
determine the
same outcome of the game $\Gamma^{\mathrm{MW}}_{2}$.\\

\begin{xx}
Let density operator $\rho_{\mathrm{in}}$ of a state
(\ref{generalinitialstate}) be given. Then the state
$\rho'_{\mathrm{fin}}$ after the third step of procedure
 can be expressed as:
\begin{align}
\rho'_{\mathrm{fin}} &=
\sigma^2_{\kappa_{2}}M_{0}\rho_{\kappa_{1}}
M_{0}\sigma^2_{\kappa_{2}} +
\sigma^3_{\kappa_{3}}M_{1}\rho_{\kappa_{1}}
M_{1}\sigma^3_{\kappa_{3}}\notag\\ &=
M_{0}\sigma^2_{\kappa_{2}}\rho_{\kappa_{1}}\sigma^2_{\kappa_{2}}M_{0}
+
M_{1}\sigma^3_{\kappa_{3}}\rho_{\kappa_{1}}\sigma^3_{\kappa_{3}}M_{1}.
\end{align}
Since $X^{\kappa}M_{\iota} = \delta_{\kappa \iota}X^{\kappa}$,
where $\delta_{\kappa \iota}$ is the  Kronecker's delta, we
obtain:
\begin{align}
\mathrm{tr}(X\rho'_{\mathrm{fin}}) =
\mathrm{tr}(X^0\sigma^2_{\kappa_{2}}\rho_{\kappa_{1}}\sigma^2_{\kappa_{2}}
+ X^1\sigma^3_{\kappa_{3}}\rho_{\kappa_{1}}\sigma^3_{\kappa_{3}}).
\label{equation1}
\end{align}
Notice that operation $\sigma_{1}$ on the second (third) qubit of
any state (\ref{generalinitialstate}) does not influence the
measurement of outcomes $O_{10}$ and $O_{11}$ ($O_{00}$ and
$O_{01}$), because of the form of $X^{1}$ ($X^{0}$), which means
that:
\begin{align}
\mathrm{tr}(X^{\iota}\sigma^{2+\iota}_{\kappa_{2+\iota}}\rho_{\kappa_{1}}\sigma^{2+\iota}_{\kappa_{2+\iota}})
= \mathrm{tr}(X^{\iota} \sigma^2_{\kappa_{2}} \otimes
\sigma^3_{\kappa_{3}}\rho_{\kappa_{1}}\sigma^2_{\kappa_{2}}
\otimes \sigma^3_{\kappa_{3}})\quad \mbox{for} \quad \iota = 0,1.
\label{notinfluence}
\end{align}
Inserting (\ref{notinfluence}) into the formula (\ref{equation1})
we get:
\begin{align}
\mathrm{tr}\left(X\rho'_{\mathrm{fin}}\right) =
\mathrm{tr}\Biggl(\left(X^{0} +
X^{1}\right)\Biggl(\bigotimes^3_{j=1}
\sigma^j_{\kappa_{j}}\rho_{\mathrm{in}}\bigotimes^3_{j=1}
\sigma^j_{\kappa_{j}}\Biggr)\Biggr). \label{formulakoniec}
\end{align}
The right-hand side of (\ref{formulakoniec}) is equal the expected
outcome given by formula (\ref{eformulaultimatum}). Thus, the two
ways of determining the final state are outcome-equivalent.
\end{xx}
\noindent We claim that performing quantum measurement is a more
natural manner to play quantum games than observation of player's
actions taken previously - the way suggested by games played
classically. Since the result of a quantum game is determined by
the measurement outcome of the final state instead of actions
taken by players, each stage of the quantum game also ought to be
set via a quantum measurement of a current state. Moreover, when
we suppose the second player's move dependence on actions of the
first player in $\Gamma_{2}$ then it implies the same game tree as
in Figure.~\ref{figure3}a). This way, however, stands in
contradiction to the results in subsection~\ref{subsection_MW}
that tell us that the game trees must be different. Of course, if
the initial state is $|000\rangle \langle 000|$ (i.e., when game
given by (\ref{gameMW3}) boils down to a game (\ref{gamma3})),
observation of the course of the game played classically and with
the use of quantum measurement coincide.

\begin{figure}[t]
\centering
\includegraphics[angle=0, scale=0.8]{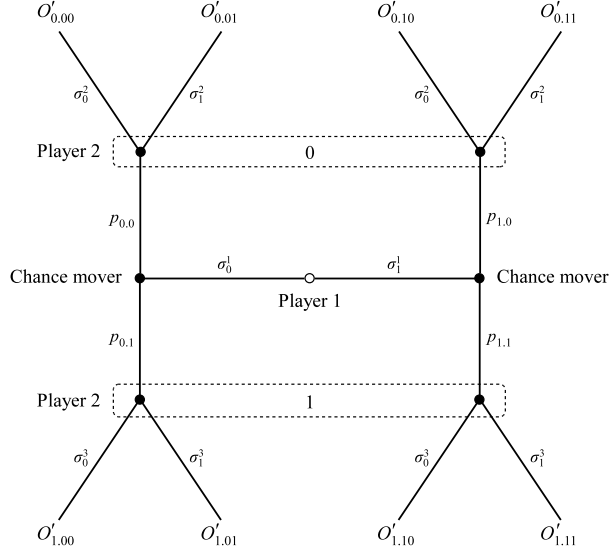}
\caption{The extensive game associated with the quantum
realization $\Gamma^{\mathrm{MW}}_{2}$.} \label{figure5}
\end{figure}

Let us study what a game tree corresponding to the game
$\Gamma^{\mathrm{MW}}_{2}$ is yielded by the above-mentioned
procedure. According to the first step, the initial history
$\emptyset$ is followed by two actions of the first player. Next,
the measurement on the first qubit is made. The two possible
measurement outcomes $\iota = 0,1$ can be identified with two
actions (following each player 1's move) of a chance mover that
are taken with probability $p_{\kappa_{1},\iota}$. Finally the
player 2 acts on $\iota +2$ qubit of the state $\rho_{\iota}$
after each history associated with the outcome~$\iota$. Therefore,
all histories followed by given outcome~$\iota$ constitutes an
information set of player 2. Such description in a form of a~game
tree is illustrated in Figure~\ref{figure5}.
The outcomes $O'_{0.\kappa_{1}{,}\kappa_{2}}$ and
$O'_{1.\kappa_{1}{,}\kappa_{3}}$ are determined by the following
equations:
\begin{align}
O'_{0.\kappa_{1}{,}\kappa_{2}} = \mathrm{tr}(X \sigma_{\kappa_{2}}
\rho_{\kappa_{1},0} \sigma_{\kappa_{2}}),\quad
O'_{1.\kappa_{1}{,}\kappa_{3}} = \mathrm{tr}(X \sigma_{\kappa_{3}}
\rho_{\kappa_{1},1} \sigma_{\kappa_{3}}). \label{ostatnierownania}
\end{align}
We have proved that the two approaches:
(\ref{finalstateultimatum}) and the sequential one to calculate
the final state are outcome-equivalent. Therefore, it should be
expected that extensive forms of $\Gamma_{2}$ and
$\Gamma^{\mathrm{MW}}_{2}$ coincide when the initial state is a
basis state. In fact, given $\rho_{\mathrm{in}} = |000\rangle
\langle 000|$ the probabilities $p_{\kappa_{1},\iota}$ are
expressed by the formula $p_{\kappa_{1},\iota} =
\delta_{\kappa_{1},\iota}$, where $\kappa_{1}, \iota \in \{0,1\}$.
Then, the available outcomes given by Eq.~(\ref{ostatnierownania})
are as follows: $O'_{0.00} = O_{00}$, $O'_{0.01} = O_{01}$,
$O'_{1.10} = O_{10}$ and $O'_{1.11} = O_{11}$. By identifying
$\sigma^1_{\kappa_{1}} \mathrel{\mathop:}= c_{\kappa_{1}}$,
$\sigma^2_{\kappa_{2}} \mathrel{\mathop:}= d_{\kappa_{2}}$,
$\sigma^3_{\kappa_{3}} \mathrel{\mathop:}= e_{\kappa_{3}}$ the
extensive game in Figure~\ref{figure5} represents game
$\Gamma_{2}$.
\section{Conclusion}
We have shown that our proposal extends the ultimatum game in the
quantum area. Although proposed scheme is suitable only for a
normal representation of the ultimatum game in which some features
of corresponding game in extensive form are lost, it passes on
valuable information about how passing to the quantum domain
influences a course of extensive games. The dominant position of
player 1, when the ultimatum game is played classically, can be
weakened in the case of playing the game via both the MW approach
and the EWL approach. Another thing worth noting is that the the
quantization significantly extends the game tree compared with
classical case. It makes the normal representation to be more
convenient way to analyze the game than the way of extensive form.

\end{document}